\documentclass[12pt]{article}
\usepackage{bm}
\usepackage{amsmath}
	\numberwithin{equation}{section}
\usepackage{amssymb}
\usepackage{graphics}
\usepackage{bbold}
\usepackage{multirow}
\usepackage{enumerate}

\begin{document}

\begin{titlepage}
  \begin{flushright}
    TIT/HEP-642 \\
    February 2015
  \end{flushright}

  \vspace{0.5cm}

  \begin{center}
    {\Large \bf
      ODE/IM correspondence and Bethe ansatz for affine Toda field equations
    }

    \lineskip .75em
    \vskip2.5cm

    {\large
      Katsushi Ito and Christopher Locke
    }

    \vskip 2.5em

    {\normalsize\it
      Department of Physics,\\

      Tokyo Institute of Technology\\
      Tokyo, 152-8551, Japan
    }

    \vskip 1.0em
    \vskip 3.0em
  \end{center}

  \begin{abstract}

    We study the linear problem associated with modified affine Toda field equation for the Langlands dual  $\hat{\mathfrak{g}}^{\vee}$, where $\hat{\mathfrak{g}}$ is an untwisted affine Lie algebra.
    The connection coefficients for the asymptotic solutions of the linear problem are found to correspond to the $Q$-functions for $\mathfrak{g}$-type quantum integrable models.
    The $\psi$-system for the solutions associated with the fundamental representations of $\mathfrak{g}$ leads to Bethe ansatz equations associated with the affine Lie algebra $\hat{\mathfrak{g}}$.
    We also study the $A^{(2)}_{2r}$ affine Toda field equation in massless limit in detail and find its Bethe ansatz equations as well as $T$-$Q$ relations.

  \end{abstract}
\end{titlepage}

\baselineskip=0.7cm

\section{Introduction} \label{sec:intro}

The ODE/IM correspondence was proposed by Dorey and Tateo in \cite{dorey1999anharmonic} where they demonstrated an interesting relationship between a Schr\"odinger-type ordinary differential equation with anharmonic potential and the conformal limit of a certain two-dimensional quantum integrable model.
It was shown that functional relations satisfied by the Stokes multipliers and spectral determinants of this ODE agree with those of the $\mathbf{Q}$-operator and transfer matrix vacuum eigenvalues for an $A_1$ type quantum integrable system in the conformal field theory limit (see also \cite{bazhanov1997integrable}).
The case where the Schr\"odinger differential equation is modified with an additional angular momentum potential was studied in \cite{bazhanov2001spectral}.
This correspondence is now just a single example of the growing number of links between classical and quantum integrable models.

The generalization of this massless ODE/IM correspondence to simple Lie algebra $A_r$ was carried out in \cite{dorey2000differential,suzuki2000functional}.
The case of other simple Lie algebras was studied in \cite{dorey2007pseudo}, where it was necessary to consider in general pseudo-differential equations.
The work of \cite{juanjuan2012polynomial} showed that the same results could be obtained by using a first order formulation that did not require introduction of a formal anti-derivative.

Lukyanov and Zamolodchikov \cite{lukyanov2010quantum} studied the ODE/IM correspondence for the massive sine(h)-Gordon model and found that spectral determinants of a modified form of the classical sinh-Gordon model coincide with the $Q$-functions of the quantum sine-Gordon model, the affine Toda field theory for algebra $A^{(1)}_1$.
This was generalized to a relation between the classical Tzitz\'eica-Bullough-Dodd equation ($A^{(2)}_2$ algebra) and the quantum Izergin-Korepin model in \cite{dorey2013bethe}, and was studied for type $A_r^{(1)}$ affine Toda theories in \cite{adamopoulou2014bethe,Ito2014600}.
In these works it was shown that connection coefficients for subdominant solutions to the linear problem associated with the affine Toda field equation correspond to the vacuum eigenvalues of $\mathbf{Q}$-operators for $\mathfrak{g}$-type quantum integrable models.
The work of \cite{Ito2014600} looked at $ABCDG$-type affine Lie algebras and found that the \mbox{(pseudo-)}ordinary differential equation associated with $\hat{\mathfrak{g}}^\vee$ affine Toda field equation was the same as that of \cite{dorey2007pseudo} for simple Lie algebra $\mathfrak{g}$ after taking the conformal limit.

While the work of \cite{lukyanov2010quantum,dorey2013bethe} used a functional relation on the subdominant solution to the linear problem to obtain Bethe ansatz equations satisfied by the $Q$-function, the connection to the previously studied $\psi$-systems was not manifest.
The $\psi$-system, a set of functional relations among uniquely defined solutions $\psi^{(a)}$ to a \mbox{(pseudo-)}ODE for $a=1,\ldots,\text{rank}(\mathfrak{g})$, was found in \cite{dorey2007pseudo} (see also \cite{juanjuan2012polynomial}).
These $\psi$-systems are similar to the Pl\"ucker type relations, and using these relations they were able to derive the Bethe ansatz equations satisfied by the $Q$-functions which corresponded to the $Q$-function of a conformal vertex model associated to $\mathfrak{g}$.

In this paper we investigate the $\psi$-system of \cite{dorey2007pseudo,juanjuan2012polynomial} and show how it also holds in the massive case for subdominant solutions to the linear problem associated to a modified affine Toda field equation for affine Lie algebra $\hat{\mathfrak{g}}^\vee$, where $\hat{\mathfrak{g}}$ is an untwisted affine algebra.
The case of $A^{(2)}_{2r}$ is unique in that it is non-simply laced yet its Langlands dual is equal to itself.
Furthermore, the correspondence in \cite{Ito2014600} links massive theories associated to the Langlands dual affine algebra $\hat{\mathfrak{g}}^\vee$ to conformal quantum theories associated with $\mathfrak{g}$ in the massless limit, so it is interesting to understand $A^{(2)}_{2r}$ which does not fit into this scheme in more detail.
To investigate the meaning in this case we also propose a new $\psi$-system for $A^{(2)}_{2r}$ and give evidence for it by studying the spectral determinant of the ordinary differential equation associated with the linear problem and find its $T$-$Q$ relations and the Bethe ansatz equations satisfied by $Q$.
The case of untwisted non-simply laced affine Lie algebras remains elusive at the moment.

The flow of this paper is as follows.
In section \ref{sec:toda} we introduce the modified form of the classical affine Toda field equation used in this paper and its linear form.
This section's main purpose is to introduce some special solutions to the linear problem determined by their asymptotic behavior near the irregular singularity at $z=\infty$ and the regular singularity at $z=0$.
Section \ref{sec:psi} introduces the $\psi$-system functional relations satisfied by uniquely determined subdominant solutions to the linear problem $\Psi^{(a)}$.
These massive $\psi$-systems serve as the fulcrum of this work, linking the classical affine Toda differential equations with $Q$-functions corresponding to some massive quantum integrable model.
Finally section \ref{sec:BAE} uses the special solutions of section \ref{sec:toda} and the functional relations of section \ref{sec:psi} to give relations satisfied by the connection coefficients $Q$ that are the same as Bethe ansatz equations for associated quantum integrable models.

\section{Affine Toda field equations} \label{sec:toda}

In this section we will first summarize the Lie algebra conventions used in this paper.
We then introduce the modified affine Toda field equation, including its linear form, and study special solutions defined by their asymptotic behaviors.

\subsection{Lie algebra preliminaries} \label{subsec:lie}

A rank $r$ Lie algebra $\mathfrak{g}$ has generators in $\{E_\alpha, H^i\}$ where $\alpha \in \Delta$ (the set of roots) and $i = 1, \ldots, r$.
The commutation relations satisfied by these generators are \cite{francesco1997conformal}
\begin{align}
  [H^i, H^j] &= 0 \,, \\
  [H^i, E_{\alpha}] &= \alpha^i \ E_{\alpha} \,, \\
  [ E_{\alpha} , E_{\beta} ] &= \left\{
    \begin{array}{cc}
      N_{\alpha,\beta} \ E_{\alpha + \beta} \quad & \text{for} \; \alpha + \beta \in \Delta \\
      \alpha^\vee \cdot H \quad & \text{for} \; \alpha + \beta = 0 \\
      0 \quad & \text{otherwise}
    \end{array}
  \right. \,,
\end{align}
where $\alpha \cdot H = \sum_{i=1}^r \alpha^i H^i$, $\alpha^2 = \sum_{i=1}^r \alpha^i \alpha^i$, $\alpha^\vee = 2\alpha / \alpha^2$ is the coroot of $\alpha$ and $N_{\alpha,\beta}$ are structure constants.
Lie algebra $\mathfrak{g}$ has fundamental weights $\omega_a$ and simple roots $\alpha_a$ where $a=1,\ldots,r$ and $\alpha_a^\vee \cdot \omega_b = \delta_{a,b}$.
The Cartan matrix is defined to be $A_{ab} = \alpha_a \cdot \alpha^\vee_b$.
We normalize the roots so that the long root has length $2$.

Let $\hat{\mathfrak{g}}$ denote the affine Lie algebra of $\mathfrak{g}$.
Its extended Dynkin diagram is obtained from that of $\mathfrak{g}$ by adding the root $\alpha_0 = -\theta$, where $\theta$ is the highest root.
The (dual) Coxeter labels $n_a$ ($n^\vee_a$) are integers satisfying $0 = \sum_{a=0}^r n_a \alpha_a = \sum_{a=0}^r n^\vee_a \alpha^\vee_a$ and $n_0^\vee = 1$.
The (dual) Coxeter number $h$ ($h^\vee$) is the sum of the (dual) Coxeter labels, and the (co)Weyl vector $\rho$ ($\rho^\vee$) is the sum of the (co)fundamental weights.
$\hat{\mathfrak{g}}^{\vee}$ denotes the Langlands dual of $\hat{\mathfrak{g}}$, whose simple roots are $\alpha_{a}^{\vee}$.
The simply-laced affine Lie algebras $A^{(1)}_r$, $D^{(1)}_r$, and $E^{(1)}_r$ are self-dual, whereas the non simply-laced cases obey $(B^{(1)}_r)^\vee = A^{(2)}_{2r-1}$, $(C^{(1)}_r)^\vee = D^{(2)}_{r+1}$, $(F^{(1)}_4)^\vee = E^{(2)}_{6}$, $(G^{(1)}_2)^\vee = D^{(3)}_4$, and $(A^{(2)}_{2r})^\vee = A^{(2)}_{2r}$.

\subsection{Modified affine Toda field equation} \label{subsec:mToda}

First we will define the two-dimensional affine Toda field equation associated with $\hat{\mathfrak{g}}$.
The theory is defined on the complex plane using coordinates
\begin{gather}
  z = \frac{1}{2} (x^0 + i x^1) \,, \quad \bar{z} = \frac{1}{2} (x^0 - i x^1) \,, \quad z = \rho e^{i\theta} ,
\end{gather}
where $\rho$ and $\theta$ are polar coordinates.
The equation of motion for the two-dimensional modified affine Toda equation studied here is\footnote{Note that we choose the sign of the kinetic term in equation \eqref{eq:mtodaeq} to be opposite that of \cite{Ito2014600}; this agrees with the conventions of \cite{lukyanov2010quantum,dorey2013bethe} and makes large $z$ asymptotic analysis a little nicer.}
\begin{gather}
  \beta \partial\bar{\partial}\phi - m^2 \left[\sum_{i=1}^{r} n_i \alpha_i e^{\beta \alpha_i \cdot \phi} + p(z)\bar{p}(\bar{z}) n_0 \alpha_0 e^{\beta \alpha_0 \cdot \phi} \right] = 0 \,, \label{eq:mtodaeq}
\end{gather}
where $\phi$ is a vector of $r$ scalar fields, $\beta$ a dimensionless coupling constant and $m$ a mass parameter.
The conformal factor $p(z)$ here is chosen to have the form (see \cite{lukyanov2010quantum,dorey2013bethe})
\begin{gather}
  p(z,s) = z^{hM} - s^{hM} \,. \label{eq:p}
\end{gather}
Here we take $M > 0$ real positive and $s$ complex.

Equation \eqref{eq:mtodaeq} can be written as a zero curvature condition, $\mathrm{d} \mathbf{A} + \mathbf{A} \wedge \mathbf{A} = 0$, where $\mathbf{A} = A \ \mathrm{d}z + \bar{A} \ \mathrm{d}\bar{z}$ is the $\mathfrak{g}$-valued one form with
\begin{gather}
  A = \frac{\beta}{2} \partial \phi \cdot H + m e^{\lambda} \left[ \sum_{i=1}^r \sqrt{n_i^\vee} e^{\beta \alpha_i \cdot \phi / 2} E_{\alpha_i} + p(z) \sqrt{n_0^\vee} e^{\beta \alpha_0 \cdot \phi / 2} E_{\alpha_0} \right] \label{eq:A_conn} , \\
  \bar{A} = -\frac{\beta}{2} \bar\partial \phi \cdot H + m e^{-\lambda} \left[ \sum_{i=1}^r \sqrt{n_i^\vee} e^{\beta \alpha_i \cdot \phi / 2} E_{-\alpha_i} + \bar{p}(\bar{z}) \sqrt{n_0^\vee} e^{\beta \alpha_0 \cdot \phi / 2} E_{-\alpha_0} \right] \label{eq:A_bar_conn} .
\end{gather}
Here we introduced the spectral parameter $\lambda$.
This zero curvature condition can equivalently be written as a first order linear problem defined on some finite dimensional $\mathfrak{g}$-module,
\begin{gather}
  (\mathrm{d} + \mathbf{A}) \Psi = 0 \label{eq:linear_problem} \,.
\end{gather}
Such connections can be changed through an arbitrary gauge transformation of the form
\begin{gather}
  \tilde{\mathbf{A}} = U \mathbf{A} U^{-1} + U \mathrm{d} U^{-1} , \quad \tilde{\Psi} = U \Psi . \label{eq:gauge}
\end{gather}
This leaves the zero curvature condition and linear problem unchanged, and will be used to put the connection into various convenient forms.

\subsection{Asymptotic behavior}

Now we will look at the asymptotic behavior of solutions to the modified affine Toda field equation and its linear problem.

First, following \cite{lukyanov2010quantum,dorey2013bethe,adamopoulou2014bethe,Ito2014600} we consider a special family of solutions to the equation of motion \eqref{eq:mtodaeq} $\phi(\rho,\theta)$ with the following properties:

\begin{enumerate}[{\bf (i)}]
  \item
    Consistent with the choice of $p(z)$ in \eqref{eq:p}, $\phi(\rho,\theta)$ should have periodicity:
    \begin{gather}
      \phi\left(\rho, \theta + \tfrac{2\pi}{hM}\right) = \phi(\rho,\theta) .
    \end{gather}

  \item
    The field $\phi(\rho,\theta)$ is real-valued for real $\rho$ and $\theta$ (i.e.\ when $\bar{z}$ is identified as the complex conjugate of $z$), and finite everywhere except at the apex $\rho = 0$.

  \item
    For large $\rho$, $\phi(\rho,\theta)$ has logarithmic divergence,
    \begin{gather}
      \phi(\rho, \theta) = \frac{2 M \rho^\vee}{\beta} \log(\rho) + \mathcal{O}(1) \quad \text{as} \;\; \rho \rightarrow \infty .
    \end{gather}

  \item
    For $\rho \sim 0$, the field $\phi(\rho,\theta)$ diverges logarithmically,
    \begin{gather}
      \phi(\rho,\theta) = 2g \log(\rho) + \mathcal{O}(1) \quad \text{for} \;\; \beta \alpha_a \cdot g + 1 > 0 , \;\; a = 0, 1, \ldots, r .
    \end{gather}
    Here $g$ is an $r$ component vector that parameterizes the behavior of $\phi(\rho,\theta)$ near $0$.
\end{enumerate}

The periodicity condition naturally leads one to define the following transformation under which both the equation of motion and linear problem are unchanged for integer $k$,
\begin{align}
  \hat{\Omega}_k : \; \left\{
  \begin{matrix}
    & z \rightarrow z \ e^{2\pi k i / hM} \\
    & s \rightarrow s \ e^{2\pi k i / hM} \\
    & \lambda \rightarrow \lambda - \tfrac{2\pi k i}{hM}
  \end{matrix} \right. \;. \label{eq:mToda_sym}
\end{align}
Functions that are rotated by this transformation are said to be $k$-Symanzik rotated, and will often be denoted with a subscript as follows,
\begin{gather}
  \hat{\Omega}_k f(z,\bar{z}) = f_k(z,\bar{z}) \,.
\end{gather}
The linear problem also has another symmetry,
\begin{gather}
  \hat{\Pi} : \; \left\{
  \begin{matrix}
    & \lambda & \rightarrow & \lambda - \frac{2 \pi i}{h} \\
    & \mathbf{A} & \rightarrow & S \mathbf{A} S^{-1} \\
    & \Psi & \rightarrow & S \Psi
  \end{matrix} \right. \,, \quad\quad S = \exp \left(\frac{2\pi i}{h} \rho^\vee \cdot H \right) \,.
\end{gather}
This symmetry follows naturally by noticing that under $\hat{\Pi}$, $E_{\alpha_i}$ transforms as
\begin{gather}
  E_{\alpha_i} \rightarrow S E_{\alpha_i} S^{-1} = e^{2\pi i / h} E_{\alpha_i} \,, \;\; \text{for $i = 0, 1, \ldots r$} \,.
\end{gather}

For the following we will consider the linear problem \eqref{eq:linear_problem} in $\mathfrak{g}$-module $V^{(a)}$ where the representation of this module has highest weight $\omega_a$ and dimension $\prod_{\alpha>0} \tfrac{(\omega_a + \rho) \cdot \alpha}{\rho \cdot \alpha}$ \cite{francesco1997conformal} where $\rho$ is the Weyl vector, half the sum of the positive roots.
The vector space $V^{(a)}$ has a basis $\mathbf{e}_i^{(a)}$ for $i=1,\ldots,\dim(V^{(a)})$, where each basis vector is naturally associated with a weight $h^{(a)}_i$ such that $H^i \mathbf{e}^{(a)}_j = (h^{(a)}_j)^i \mathbf{e}_j^{(a)}$, and the basis vector associated with the highest weight $\omega_a = h^{(a)}_1$ is by convention $\mathbf{e}^{(a)}_1$.

In this work, we will be interested in the unique solution $\Psi^{(a)}$ in module $V^{(a)}$ that is \emph{subdominant}, that is, the solution that decays fastest along the positive real axis.
To find this subdominant solution it is useful to take a gauge transformation \eqref{eq:gauge} that puts either the holomorphic or anti-holomorphic connection into a nice form with no exponentials where $U$ is respectively
\begin{gather}
	U_A = z^{M \rho^\vee \cdot H} e^{-\beta \phi \cdot H / 2} \,, \quad U_{\bar{A}} = \bar{z}^{-M \rho^\vee \cdot H} e^{\beta \phi \cdot H / 2} .
\end{gather}
In the large $z$ limit, $\phi(z,\bar{z}) \sim \tfrac{M \rho^\vee}{\beta} \log(z\bar{z})$ and $p(z) \sim z^{hM}$, and the connections become
\begin{gather}
	\tilde{A} = m e^\lambda z^M \Lambda_+ , \quad \tilde{\bar{A}} = m e^{-\lambda} \bar{z}^M \Lambda_- , \\
    \Lambda_\pm = \sqrt{n_0^\vee} E_{\pm \alpha_0} + \sum_{i=1}^r \sqrt{n_i^\vee} E_{\pm \alpha_i} . \label{eq:Lambda}
\end{gather}

Now the subdominant solution is found to be, through consideration of the holomorphic and anti-holomorphic linear problems separately and then shifting back to the original $\Psi = U^{-1} \tilde{\Psi}$,
\begin{gather}
  \Psi^{(a)} = g(\bar{z}) \exp\left( -\mu_+^{(a)} \frac{z^{M+1}}{M+1} m e^\lambda\right) (z / \bar{z})^{-M \rho^\vee \cdot H / 2} \boldsymbol{\mu}_+^{(a)} , \label{eq:psia_holo}\\
  \Psi^{(a)} = f(z) \exp\left( -\mu_-^{(a)} \frac{\bar{z}^{M+1}}{M+1} m e^{-\lambda}\right) (z / \bar{z})^{-M \rho^\vee \cdot H / 2} \boldsymbol{\mu}_-^{(a)} , \label{eq:psia_antiholo}
\end{gather}
where $\mu^{(a)}_\pm$ and $\boldsymbol{\mu}^{(a)}_\pm$ are the eigenvalues of $\Lambda_\pm$ with the largest real part and its eigenvector in module $V^{(a)}$.
This eigenvalue is distinct, and furthermore since the representations can be chosen such that $E_{\alpha}^\top = E_{-\alpha}$, we have $\Lambda_- = (\Lambda_+)^\top$ and the two eigenvalues and eigenvectors are the same.

Finally, after setting the $\Psi^{(a)}$ from \eqref{eq:psia_holo} and \eqref{eq:psia_antiholo} to be equal, $f$ and $g$ are fixed within a constant giving
\begin{gather}
	\Psi^{(a)} = \exp\left( -2 \mu^{(a)} \frac{\rho^
    {M+1}}{M+1} m \cosh\left(\lambda + i \theta (M+1)\right) \right) e^{-i \theta M \rho^\vee \cdot H} \boldsymbol{\mu}^{(a)}. \label{eq:psi_a}
\end{gather}
Applying $\hat{\Omega}_k$ to this for any real number $k$ gives the $k$-Symanzik rotated solution
\begin{gather}
  \Psi^{(a)}_k = \exp\left( -2 \mu^{(a)} \frac{\rho^{M+1}}{M+1} m \cosh\left(\lambda + i\theta(M+1) + \tfrac{2\pi i k}{h}\right) \right) e^{-i(\theta M + \tfrac{2\pi k}{h}) \rho^\vee \cdot H} \boldsymbol{\mu}^{(a)} . \label{eq:psi_a_k}
\end{gather}
Note that a $\hat{\Pi}$ transformation applied to $\Psi^{(a)}$ gives the same large-$\rho$ behavior as $\Psi^{(a)}_{-1}$.
$\Psi^{(a)}_k$ is the subdominant solution in the Stokes sector
\begin{gather}
  \mathcal{S}_k \; : \quad \left| \theta + \frac{2\pi k}{h(M+1)} \right| < \frac{\pi}{h(M+1)} \,.
\end{gather}

A basis of solutions to the linear problem defined by their behavior around $0$ can also be defined by setting all components of $\Psi$ to zero around $\rho=0$ except for a single component $\mathbf{e}^{(a)}_i$ (see \cite{lukyanov2010quantum,dorey2013bethe}).
By considering the holomorphic and anti-holomorphic linear problem it can be shown that such a solution $\mathcal{X}_i^{(a)}$ must satisfy
\begin{gather}
  \mathcal{X}_i^{(a)} = e^{-(\lambda + i\theta) \beta g \cdot h^{(a)}_i} \mathbf{e}^{(a)}_i + \mathcal{O}(\rho) \;\; \text{as} \;\; \rho \rightarrow 0 \,, \label{eq:small_asymp}
\end{gather}
where the overall constant's dependence on $\lambda$ was fixed by requiring that this solution is invariant under $\hat{\Omega}_k$.
Note however that the $\Psi^{(a)}$ solutions do not display this invariance under $\hat{\Omega}_k$.

Since $\mathcal{X}^{(a)}_i$ form a basis of solutions to the linear problem, the subdominant solution $\Psi^{(a)}$ can be expanded as
\begin{gather}
  \Psi^{(a)}(z,\bar{z} | \lambda,g) = \sum_{i=1}^{\dim(V^{(a)})} Q^{(a)}_i (\lambda, g) \ \mathcal{X}^{(a)}_i (z,\bar{z} | \lambda, g) . \label{eq:psi_q_exp}
\end{gather}
These coefficients $Q^{(a)}_i(\lambda,g)$ are radial spectral determinants that are only vanishing for values of the spectral parameter $\lambda$ in which there exists a solution that decays exponentially like \eqref{eq:psi_a} for large $\rho$, and has the coefficient proportional to $\mathbf{e}^{(a)}_i$ go to zero as $\rho \rightarrow 0$.
From the above observation that $\hat{\Omega}_1 \hat{\Pi} \Psi^{(a)} = \Psi^{(a)}$ and the asymptotic form of $\mathcal{X}^{(a)}_i$, one can derive a quasi-periodic condition for these $Q$-functions,
\begin{gather}
  Q^{(a)}_i\left(\lambda - \tfrac{2\pi i}{hM} (M+1), g \right) = \exp\left(-\tfrac{2\pi i}{h} (\rho^\vee + \beta g) \cdot h^{(a)}_i \right) Q^{(a)}_i(\lambda, g) \,.
\end{gather}

We conjecture that for affine Toda field equations with algebra $\hat{\mathfrak{g}}^{\vee}$ these $Q$-functions will correspond to the vacuum eigenvalues of $\mathbf{Q}$-operators for some massive integrable quantum field theory associated with $\hat{\mathfrak{g}}$, where the massive theories are known only for $A_1^{(1)}$ \cite{lukyanov2010quantum}, $A_2^{(2)}$ \cite{dorey2013bethe}, and for Fateev models \cite{bazhanov2014integrable}.
We will give evidence for this correspondence by showing that in the conformal limit these connection coefficients $Q$ will satisfy Bethe ansatz equations associated to vertex models with Langlands dual Lie algebra symmetry.

\section{$\psi$-system} \label{sec:psi}

The $\psi$-system \cite{dorey2007pseudo} is a set of Pl\"ucker type relations satisfied by auxiliary functions that are constructed from the subdominant solution to a (pseudo-)ODE.
The $\psi$-system was proved for $A$-type simple Lie algebras and was conjectured for all other simple Lie algebras.
In \cite{juanjuan2012polynomial}, the $\psi$-system for classical Lie algebras was derived by studying the first order system equivalent to the (pseudo-)ODE of \cite{dorey2007pseudo} and embeddings of $\mathfrak{g}$-modules.

We will study the $\psi$-systems in the context of modified affine Toda field equations with algebra $\hat{\mathfrak{g}}^\vee$ and show that the same system of functional relations holds for the massive case.
In particular it will be shown that the unique subdominant solutions $\Psi^{(a)}$ to the linear problem
\begin{gather}
  (\mathrm{d} + \mathbf{A}) \Psi^{(a)} = 0 \label{eq:linear_psia}
\end{gather}
in $\mathfrak{g}$-module $V^{(a)}$ satisfy the same $\psi$-system relations of \cite{juanjuan2012polynomial} for $\hat{\mathfrak{g}}^\vee$ when $\mathfrak{g}$ is a classical Lie algebra, and \cite{dorey2007pseudo} when $\mathfrak{g}$ is an exceptional Lie algebra.
We also find a new $\psi$-system for $A^{(2)}_{2r}$ affine Toda theories.

Let us consider an embedding of modules as explained in \cite{juanjuan2012polynomial} (see also \cite{fulton1991representation}).
In the case of $A_r^{(1)}$ there is an embedding $\iota$ which acts as
\begin{gather}
  \iota \;: \bigwedge^2 V^{(a)} \rightarrow V^{(a-1)} \otimes V^{(a+1)} \,,
\end{gather}
where the left-hand side is the exterior product of two $V^{(a)}$'s.
As consistency expects the highest weight of the left and right side modules are the same, $\omega_{a-1} + \omega_{a+1}$.
Next, the incidence matrix $B_{ab}$ is related to the Cartan matrix as
\begin{gather}
  B_{ab} = 2 \delta_{ab} - A_{ab} \,. \label{eq:B_def}
\end{gather}
Denoting the highest weight of $V^{(a)}$ as $h_1^{(a)}$ and the next highest weight $h_2^{(a)} = h^{(a)}_1 - \alpha_a$, then the sum satisfies
\begin{align}
  h^{(a)}_1 + h^{(a)}_2 &= \sum_{b=1}^r B_{ab} \, \omega_{b} \,.
\end{align}
So for general $\hat{\mathfrak{g}}$ there is an embedding $\iota$ which acts as
\begin{align}
  \iota : \bigwedge^2 V^{(a)} \rightarrow \bigotimes_{b=1}^r \left(V^{(b)}\right)^{B_{ab}} .
\end{align}

\subsection{Simply laced cases}

First we explain the origin of the $\psi$-system functional relations satisfied by the subdominant solutions $\Psi^{(a)}$ for the simply-laced cases.
For the $A^{(1)}_r$ case this is found to be
\begin{gather}
  \iota\left(\Psi^{(a)}_{-1/2} \wedge \Psi^{(a)}_{1/2}\right) = \Psi^{(a-1)} \otimes \Psi^{(a+1)} , \label{eq:a_psi_system}
\end{gather}
where $\Psi^{(0)} \equiv 1$, $\Psi^{(r+1)} \equiv 1$, and $\iota$ is the above embedding of modules.
The asymptotic behavior of $\Psi^{(a)}$ for large $\rho$ are determined from the eigenvalues $\mu^{(a)}$,
\begin{gather}
  \mu^{(a)} = \sin(\tfrac{\pi a}{h}) / \sin(\tfrac{\pi}{h}) \,, \label{eq:a_eigen}
\end{gather}
where $\mu^{(1)} = 1$.
Substititing these into \eqref{eq:psi_a_k} shows that both sides of \eqref{eq:a_psi_system} have the same asymptotics, and since both sides are unique subdominant solutions in their respective modules, both sides of this $\psi$-system are equal.
These eigenvalues have been explicitly checked up to rank $7$.

In general, the $\mu^{(a)}$ eigenvalues for simply-laced cases satisfy the equations
\begin{gather}
  2 \cos\left(\pi / h\right) \mu^{(a)} = \sum_{b=1}^r B_{ab} \, \mu^{(b)} , \label{eq:simple_mu}
\end{gather}
where the matrix $B_{ab}$ is defined in \eqref{eq:B_def}.
Note that for $D^{(1)}_r$ the eigenvalue with largest real part of $\Lambda_+$ in module $V^{(1)}$ is $\mu^{(1)} = \sqrt{2}$.
This is different from \cite{juanjuan2012polynomial} due to the presence of $\sqrt{n^\vee_i}$ coefficients in the definition of $\Lambda_+$ here (see \ref{eq:Lambda}), but the ratios of the $\mu^{(a)}$ are the same.
$\mu^{(a)}$ has been verified to satisfy \eqref{eq:simple_mu} for many cases.\footnote{
  $\mu^{(a)}$ were checked for the $D$-case by explicitly constructing the antisymmetric and spinor representations for up to rank $6$.
  For $E^{(1)}_6$ this was checked by explicitly constructing the antisymmetric representations $V^{(2)}$ and $V^{(3)}$ from $V^{(1)}$, $V^{(4)}$ as an antisymmetric representation of $V^{(5)}$, and the adjoint representation $V^{(6)}$ in the standard way, and then computing the largest eigenvalue of $\Lambda_+$.
  Due to computational limits, for $E^{(1)}_7$ we only verified the eigenvalues for $V^{(a)}$ where $a=1,2,5,6$, and for $E^{(1)}_8$ when $a=1,2$.
}

The eigenvalues \eqref{eq:simple_mu} imply that the $\psi$-system for simply laced cases is
\begin{gather}
  \iota\left( \Psi^{(a)}_{-1/2} \wedge \Psi^{(a)}_{1/2} \right) = \bigotimes_{b=1}^r \left( \Psi^{(b)} \right)^{B_{ab}} .
\end{gather}
When considering non-simply laced cases, there are difficulties that arise for $\hat{\mathfrak{g}} = B_r^{(1)}$, $C_r^{(1)}$, $F_4^{(1)}$, and $G_2^{(1)}$ with deriving a Bethe ansatz equation that has only simple poles, so we will not consider these untwisted non-simply laced cases in this work.

\subsection{Twisted cases}

By explicit calculation of the eigenvalues $\mu^{(a)}$ for the twisted affine Lie algebras, the $\psi$-systems are found to be:
\begin{align}
  (B^{(1)}_r)^\vee = A^{(2)}_{2r-1} :\quad
    & \iota\left(\Psi^{(a)}_{-1/2} \wedge \Psi^{(a)}_{1/2}\right) = \Psi^{(a-1)} \otimes \Psi^{(a+1)} \ \text{ for $a=1,\ldots,r-1$} , \notag\\
    & \iota\left(\Psi^{(r)}_{-1/4} \wedge \Psi^{(r)}_{1/4}\right) = \Psi^{(r-1)}_{-1/4} \otimes \Psi^{(r-1)}_{1/4} . \label{eq:psi_bvee}
\end{align}
\begin{align}
  (C^{(1)}_r)^\vee = D^{(2)}_{r+1} :\quad
    & \iota\left(\Psi^{(a)}_{-1/4} \wedge \Psi^{(a)}_{1/4}\right) = \Psi^{(a-1)} \otimes \Psi^{(a+1)} \ \text{ for $a=1,\ldots,r-2$} , \notag\\
    & \iota\left(\Psi^{(r-1)}_{-1/4} \wedge \Psi^{(r-1)}_{1/4}\right) = \Psi^{(r-2)} \otimes \Psi^{(r)}_{-1/4} \otimes \Psi^{(r)}_{1/4} , \notag\\
    & \iota\left(\Psi^{(r)}_{-1/2} \wedge \Psi^{(r)}_{1/2}\right) = \Psi^{(r-1)} . \label{eq:psi_cvee}
\end{align}
\begin{align}
  (F_4^{(1)})^\vee = E^{(2)}_6 :\quad
    & \iota\left(\Psi^{(1)}_{-1/2} \wedge \Psi^{(1)}_{1/2}\right) = \Psi^{(2)} , \notag\\
    & \iota\left(\Psi^{(2)}_{-1/2} \wedge \Psi^{(2)}_{1/2}\right) = \Psi^{(1)} \otimes \Psi^{(3)} , \notag\\
    & \iota\left(\Psi^{(3)}_{-1/4} \wedge \Psi^{(3)}_{1/4}\right) = \Psi^{(2)} \otimes \Psi^{(4)}_{-1/4} \otimes \Psi^{(4)}_{1/4} , \notag\\
    & \iota\left(\Psi^{(4)}_{-1/4} \wedge \Psi^{(4)}_{1/4}\right) = \Psi^{(3)} . \label{eq:psi_fvee}
\end{align}
\begin{align}
  (G^{(1)}_2)^\vee = D^{(3)}_{4} :\quad
    & \iota\left(\Psi^{(1)}_{1/2} \wedge \Psi^{(1)}_{1/2}\right) = \Psi^{(2)} ,\notag\\
    & \iota\left(\Psi^{(2)}_{1/6} \wedge \Psi^{(2)}_{1/6}\right) = \Psi^{(1)}_{-2/6} \otimes \Psi^{(1)}_0 \otimes \Psi^{(1)}_{2/6} . \label{eq:psi_gvee}
\end{align}
In \cite{dorey2007pseudo} the auxiliary functions $\psi^{(a)}$ that satisfy these relations (with Wronskian in place of wedge product) were constructed from a single subdominant solution to a \mbox{(pseudo-)}ODE, whereas here $\Psi^{(a)}$ are naturally associated to node $a$ of the Dynkin diagram since they are the subdominant solutions to the linear problem in module $V^{(a)}$.
These $\psi$-systems for affine algebra $\hat{\mathfrak{g}}^\vee$ are the same as those of \cite{dorey2007pseudo} (and \cite{juanjuan2012polynomial} for the non-exceptional cases) for simple Lie algebra $\mathfrak{g}$.

For the twisted affine Lie algebras, we need to introduce different Symanzik rotations for nodes with different root lengths on the left hand side, and some factors on the right hand side must be Symanzik rotated in order to recover the results of \cite{dorey2007pseudo} in the massless limit.
The Symanzik rotations are chosen such that both sides have the same subdominant behavior in their respective modules, and so that both sides satisfy the same linear problem.
For instance consider the first equation of \eqref{eq:psi_cvee}.
From \cite{juanjuan2012polynomial} it is known that $V^{(2)} = \bigwedge^2 V^{(1)}_{1/4}$, where the subscript $1/4$ here indicates that the action of the linear operator $(d + \mathbf{A})$ is Symanzik rotated when operating on the module to $(d + \mathbf{A})_{1/4}$.
The explicitly calculated $\mu^{(a)}$ along with this leave the above $\psi$-system as the only possibility.

\subsubsection{$A_{2r}^{(2)}$-type case}

Since there is no simple Lie algebra $X_r$ such that $(X_r^{(1)})^\vee = A^{(2)}_{2r}$, this case does not fall under the identification above of massive $\psi$-system with algebra $\hat{\mathfrak{g}}^\vee$ with massless $\psi$-system of algebra $\mathfrak{g}$.
Nevertheless, a study of the eigenvalues of $\Lambda^+$ for this case show that the $\psi$-system that $\Psi^{(a)}$ satsifies is\footnote{
The definition of twisting used is from \cite{kac1994infinite}, where the role of $\alpha$ and $\alpha^\vee$ are swapped compared with \cite{Ito2014600};
here $\alpha_0^2 = \tfrac{1}{2}$, $\alpha_i^2=1$, and $\alpha_r^2 = 2$, while $n_0^\vee=1$ and $n_i^\vee=2$.
}
\begin{gather}
  \iota\left(\Psi^{(a)}_{-1/2} \wedge \Psi^{(a)}_{ 1/2}\right) = \Psi^{(a-1)} \otimes \Psi^{(a+1)} , \notag\\
  \iota\left(\Psi^{(r)}_{-1/2} \wedge \Psi^{(r)}_{ 1/2}\right) = \Psi^{(r-1)} \otimes \Psi^{(r)} . \label{eq:A2r_psi}
\end{gather}
When $r=1$ this is the same functional relation as equation (4.77) in \cite{dorey2013bethe} for the case of the Tzitz\'eica-Bullough-Dodd model.

\section{Bethe ansatz equations} \label{sec:BAE}

Using the above $\psi$-systems, it is now possible to derive functional relations for the $Q$-functions defined in equation \eqref{eq:psi_q_exp} that will correspond to Bethe ansatz equations.
We will verify that for modified affine Toda field equation with algebra $\hat{\mathfrak{g}}^\vee$, when taking the conformal limit the $Q$-functions satisfy Bethe ansatz equations associated with $\mathfrak{g}$ found in the context of the massless ODE/IM correspondence \cite{dorey2007pseudo}.

The conformal limit for modified affine Toda field equation with algebra $\hat{\mathfrak{g}}$ discussed in section \ref{sec:toda} is reached using the following definitions,
\begin{equation}
\begin{gathered}
  x = (me^{\lambda})^{1/(M+1)} z, \;\; E = s^{hM} (me^\lambda)^{hM/(M+1)} \,,\\
  \tilde{x} = (me^{-\lambda})^{1/(M+1)} \bar{z}, \;\; \tilde{E} = s^{hM} (me^{-\lambda})^{hM/(M+1)} \,.
\end{gathered} \label{eq:conf_trans}
\end{equation}
First take the light-cone limit $\bar{z} \rightarrow 0$, then send $\lambda \rightarrow \infty$ and $z, s \rightarrow 0$ while keeping $x$ and $E$ finite.
The term proportional to $p(z) \bar{p}(\bar{z})$ then drops out of the equation of motion for modified affine Toda field theory \eqref{eq:mtodaeq} with algebra $\hat{\mathfrak{g}}$ and it becomes the conformal Toda field theory associated with $\mathfrak{g}$.

Next, in the conformal limit the top component (the part proportional to the basis vector with highest weight, $\mathbf{e}^{(a)}_1$) of the expansion \eqref{eq:psi_q_exp} of $\Psi^{(a)}$ becomes
\begin{gather}
	\psi^{(a)}(x,E) = Q^{(a)}(E)\,\chi^{(a)}_1(x,E) + \tilde{Q}^{(a)}(E)\,\chi^{(a)}_2(x,E) + \cdots \,, \label{eq:psi_small}
\end{gather}
where $\psi^{(a)}$ and $\chi^{(a)}_i$ are the top components of the conformal limit of the vector wavefunctions $\Psi^{(a)}$ and $\mathcal{X}^{(a)}_i$ respectively.
The small $x$ asymptotic behavior of $\chi_i^{(a)}$ ($i=1, \ldots, \dim V^{(a)}$) can be determined through the linear problem and $\mathcal{X}^{(a)}_i$, and is
\begin{gather}
  \chi^{(a)}_i(x,E) \sim x^{\lambda^{(a)}_i} , \\
  \lambda_i^{(a)} = \rho^\vee \cdot (\omega_a - h_i^{(a)}) - \beta h^{(a)}_i \cdot g \,.
\end{gather}
Since $\lambda^{(a)}_2 - \lambda^{(a)}_1 = 1 + \beta \alpha_a \cdot g > 0$, and similarly these exponents $\lambda^{(a)}_i$ identically increase going down any link in the chain of weights, $\chi^{(a)}_1$ and $\chi^{(a)}_2$ are the most and second most dominant terms near $x=0$ in \eqref{eq:psi_small}.

Now we will demonstrate the procedure for obtaining the Bethe ansatz equations for the case of $A^{(1)}_r$.
Substituting equation \eqref{eq:psi_small} into the corresponding $\psi$-system and picking off the topmost component on both sides gives,
\begin{gather}
  \left( \omega^{-\tfrac{1}{2}(\lambda^{(a)}_1 - \lambda^{(a)}_2)} Q_{-1/2}^{(a)} \tilde{Q}_{1/2}^{(a)} - \omega^{\tfrac{1}{2}(\lambda^{(a)}_1 - \lambda^{(a)}_2)} Q_{1/2}^{(a)} \tilde{Q}_{-1/2}^{(a)} \right) W[\chi^{(a)}_0, \chi^{(a)}_1] = Q^{(a-1)} Q^{(a+1)} \chi^{(a-1)} \chi^{(a+1)} , \label{eq:QQtilde}
\end{gather}
where $\omega \equiv e^{2\pi i / h(M+1)}$ and $h$ is the Coxeter number for affine Lie algebra $\hat{\mathfrak{g}}$.
Under a $k$-Symanzik rotation in the conformal limit $Q^{(a)}_k(E)$ becomes $Q^{(a)}(\omega^{hMk} E)$.
By noting that $\lambda^{(a)}_1 + \lambda^{(a)}_2 - 1 = \lambda^{(a-1)}_1 + \lambda^{(a+1)}_1$, the $\chi$ part on both sides of \eqref{eq:QQtilde} are equal and so the following relation amongst the $Q$-functions holds,
\begin{gather}
	\omega^{-\tfrac{1}{2}(\lambda^{(a)}_1 - \lambda^{(a)}_2)} Q_{-1/2}^{(a)} \tilde{Q}_{1/2}^{(a)} - \omega^{\tfrac{1}{2}(\lambda^{(a)}_1 - \lambda^{(a)}_2)} Q_{1/2}^{(a)} \tilde{Q}_{-1/2}^{(a)} = Q^{(a-1)} Q^{(a+1)} . \label{eq:A_Q_rel}
\end{gather}
Denoting the zeros of $Q^{(a)}(E)$ as $E_k^{(a)}$, then taking a $1/2$ and $-1/2$ Symanzik rotation of \eqref{eq:A_Q_rel} evaluated at $E^{(a)}_k$ and dividing the two equations gives the functional relation
\begin{gather}
	\left. \frac{Q^{(a-1)}_{-1/2} Q^{(a)}_{1} Q^{(a+1)}_{-1/2}}{Q^{(a-1)}_{1/2} Q^{(a)}_{-1} Q^{(a+1)}_{1/2}} \right|_{E^{(a)}_k} = -\omega^{1 + \beta \alpha_a \cdot g} .
\end{gather}
These functional relations are exactly the same as the Bethe ansatz equations for $A_r$-type conformal vertex models.

This method applied to the other algebras give Bethe ansatz equations
\begin{itemize}
  \item
    $A^{(1)}_r$, $D^{(1)}_r$, $E^{(1)}_r$:
    \begin{gather}
      \left. \prod_{b=1}^r \frac{Q^{(b)}_{A_{ab} / 2}}{Q^{(b)}_{-A_{ab} / 2}} \right|_{E^{(a)}_k} = -\omega^{1 + \beta \alpha_a \cdot g} \,.
    \end{gather}

  \item
    $(B_r^{(1)})^\vee = A^{(2)}_{2r-1}$:
    \begin{align}
      \left. \frac{Q^{(a-1)}_{-1/2} Q^{(a)}_{1} Q^{(a+1)}_{-1/2}}{Q^{(a-1)}_{1/2} Q^{(a)}_{-1} Q^{(a+1)}_{1/2}} \right|_{E^{(a)}_i} &= -\omega^{1 + \beta \alpha_a \cdot g}  \ \ \text{ for $a=1,\ldots,r-1$} , \notag\\
      \left. \frac{Q^{(r-1)}_{-1/2} Q^{(r)}_{1/2}}{Q^{(r-1)}_{1/2} Q^{(r)}_{-1/2}} \right|_{E^{(r)}_i} &= -\omega^{\tfrac{1}{2}(1 + \beta \alpha_r \cdot g)} .
    \end{align}

  \item
    $(C_r^{(1)})^\vee = D^{(2)}_{r+1}$:
    \begin{align}
      \left. \frac{Q^{(a-1)}_{-1/4} Q^{(a)}_{1/2} Q^{(a+1)}_{-1/4}}{Q^{(a-1)}_{1/4} Q^{(a)}_{-1/2} Q^{(a+1)}_{1/4}} \right|_{E^{(a)}_i} &= -\omega^{\tfrac{1}{2}(1 + \beta \alpha_a \cdot g)}  \ \ \text{ for $a=1,\ldots,r-2$} , \notag\\
      \left. \frac{Q^{(r-2)}_{-1/4} Q^{(r-1)}_{1/2} Q^{(r)}_{-1/2} }{Q^{(r-2)}_{1/4} Q^{(r-1)}_{-1/2} Q^{(r)}_{1/2}} \right|_{E^{(r-1)}_i} &= -\omega^{\tfrac{1}{2}(1 + \beta \alpha_{r-1} \cdot g)} , \notag\\
      \left. \frac{Q^{(r-1)}_{-1/2} Q^{(r)}_{1}}{Q^{(r-1)}_{1/2} Q^{(r)}_{-1}} \right|_{E^{(r)}_i} &= -\omega^{1 + \beta \alpha_r \cdot g} .
    \end{align}

  \item
    $(F_4^{(1)})^\vee = E^{(2)}_6$:
    \begin{align}
        \left. \frac{Q^{(1)}_{1} Q^{(2)}_{-1/2}}{Q^{(1)}_{-1} Q^{(2)}_{1/2}} \right|_{E^{(1)}_i} = -\omega^{1 + \beta \alpha_1 \cdot g} \,,\quad & \left. \frac{Q^{(1)}_{-1/2} Q^{(2)}_{1} Q^{(3)}_{-1/2}}{Q^{(1)}_{1/2} Q^{(2)}_{-1} Q^{(3)}_{1/2}} \right|_{E^{(2)}_i} = -\omega^{1 + \beta \alpha_2 \cdot g} , \notag\\
        \left. \frac{Q^{(1)}_{-1/4} Q^{(2)}_{1/2} Q^{(3)}_{-1/2}}{Q^{(1)}_{1/4} Q^{(2)}_{-1/2} Q^{(3)}_{1/2}} \right|_{E^{(3)}_i} = -\omega^{\tfrac{1}{2}(1 + \beta \alpha_3 \cdot g)} \,,\quad & \left. \frac{Q^{(3)}_{-1/4} Q^{(4)}_{1/2}}{Q^{(3)}_{1/4} Q^{(4)}_{-1/2}} \right|_{E^{(4)}_i} = -\omega^{\tfrac{1}{2}(1 + \beta \alpha_4 \cdot g)} .
    \end{align}

  \item
    $(G_2^{(1)})^\vee = D^{(3)}_4$:
    \begin{align}
      \left. \frac{Q^{(1)}_{1} Q^{(2)}_{-1/2}}{Q^{(1)}_{-1} Q^{(2)}_{1/2}} \right|_{E^{(1)}_i} &= -\omega^{1 + \beta \alpha_1 \cdot g} , \notag\\
      \left. \frac{Q^{(1)}_{-1/2} Q^{(2)}_{2/6}}{Q^{(1)}_{1/2} Q^{(2)}_{-2/6}} \right|_{E^{(2)}_i} &= -\omega^{\tfrac{1}{3}(1 + \beta \alpha_2 \cdot g)} .
    \end{align}
\end{itemize}
Each of these Bethe ansatz equations for algebra $\hat{\mathfrak{g}}^\vee$ agree with those reported in \cite{dorey2007pseudo} (see also \cite{reshetikhin1987towards}) for massless cases with algebra $\mathfrak{g}$.

For the case of $A^{(2)}_{2r}$, the same procedure gives Bethe ansatz equations
\begin{equation}
  \begin{aligned}
	\left. \frac{Q^{(a-1)}_{-1/2} Q^{(a)}_{1} Q^{(a+1)}_{-1/2}}{Q^{(a-1)}_{1/2} Q^{(a)}_{-1} Q^{(a+1)}_{1/2}} \right|_{E^{(a)}_i} &= -\omega^{1 + \beta \alpha_a \cdot g}  \ \ \text{ for $a=1,\ldots,r-1$} , \\
	\left. \frac{Q^{(r-1)}_{-1/2} Q^{(r)}_{-1/2}Q^{(r)}_{1}}{Q^{(r-1)}_{1/2} Q^{(r)}_{1/2}Q^{(r)}_{-1}} \right|_{E^{(r)}_i} &= -\omega^{1 + \beta \alpha_r \cdot g} .
  \end{aligned} \label{eq:A2r_BAE}
\end{equation}
Note that for $A_2^{(2)}$, which corresponds to the Tzitz\'eica-Bullough-Dodd model discussed in \cite{dorey2013bethe}, $\alpha_r = \omega_r - \omega_{r-1}$ and \eqref{eq:A2r_BAE} reduces to their equation (4.85).
As mentioned this case has no corresponding massless theory from \cite{dorey2007pseudo,juanjuan2012polynomial}, so it is important to verify these equations.
To this end in appendix \ref{app:A2r_TQ} we derived the $T$-$Q$ relations that give rise to Bethe ansatz equations \eqref{eq:A2r_BAE} starting from an analysis of the ODE itself and not using the $\psi$-system (see \cite{Dorey:1999pv} for the $A_2^{(2)}$ case).
Furthermore, in appendix \ref{app:kuniba_suzuki} we also show how these Bethe ansatz equations can be found in the work of \cite{kuniba_suzuki_1995} which looked at Bethe ansatz equations associated with twisted quantum affine Lie algebras.

\section{Discussion} \label{sec:discussion}

In this paper we studied a classical affine Toda field theory for affine Lie algebra $\hat{\mathfrak{g}}^\vee$ that is modified by a conformal transformation.
Writing this modified affine Toda field equation in the linear form $(\mathrm{d} + \mathbf{A}) \Psi = 0$ translates the problem into a holomorphic and antiholomorphic first order matrix ordinary differential equation.
Studying the asymptotic behavior of solutions $\Psi$ to this linear problem, a unique subdominant solution $\Psi^{(a)}$ is found depending on the module $V^{(a)}$ in which the vector $\Psi$ lives.
These subdominant solutions $\Psi^{(a)}$ were then found to obey a set of functional relations, the massive $\psi$-system (see \cite{dorey2007pseudo,juanjuan2012polynomial} for massless case).
By expanding $\Psi^{(a)}$ in the basis of solutions $\mathcal{X}^{(a)}_i$, one can define $Q$-functions as the coefficient of $\mathcal{X}^{(a)}_1$ in this expansion.
Substituting this expansion then into the $\psi$-system in the conformal limit gives a set of functional relations on the $Q$-functions that is of the same form as Bethe ansatz equations associated with a $\mathfrak{g}$-type conformal quantum vertex model.
This was carried out for modified affine Toda field equations with algebra $\hat{\mathfrak{g}}^\vee$ where $\mathfrak{g}$ is a simple Lie algebra and the resulting Bethe ansatz system matched those of \cite{dorey2007pseudo,juanjuan2012polynomial} on the massless ODE/IM correspondence under the identification:
\begin{gather}
  \begin{matrix}
    \text{This paper} & & \text{Dorey et al.\ \cite{dorey2007pseudo}} \\
    (X^{(1)}_r)^\vee \quad & \Rightarrow & \quad X_r
  \end{matrix} \label{eq:massive_massless_ident}
\end{gather}
The presence of the Langlands dual affine algebra hints that the ODE/IM correspondence here could be a manifestation of Langlands duality \cite{feigin2007quantization}.

This identification under the conformal limit gives important evidence in support of our conjecture that the proposed $\psi$-systems hold for massive systems and that the ODE/IM correspondence links the classical modified affine Toda equations to a massive quantum integrable model.
Furthermore, previous work on the massive ODE/IM correspondence in this context on the modified sinh-Gordon equation \cite{lukyanov2010quantum} and $A^{(1)}_r$-type Toda theories \cite{adamopoulou2014bethe} are in agreement with this work.
$A^{(2)}_{2r}$ does not fall under the identification \eqref{eq:massive_massless_ident}, yet the results for this case also agreed with previous work on the Tzitz\'eica-Bullough-Dodd equation (the specific case of $A^{(2)}_2$) \cite{dorey2013bethe}.
This leads us to make the conjecture that the massive ODE/IM correspondence proposed here should hold for the simply-laced and twisted modified affine Toda field equations.

For future work, it would be worthwhile to study in detail the modified affine Toda field equation with algebra $B^{(1)}_r$, $C^{(1)}_r$, $F^{(1)}_4$, and $G^{(1)}_2$, which are the untwisted non simply-laced affine Lie algebras and do not fall under the identification \eqref{eq:massive_massless_ident} in the conformal limit.
Also, the massive ODE/IM correspondence was recently studied in the case of the classical modified sinh-Gordon equation for a choice of $p(z)$ defined on the 3-punctured Riemann sphere, and was found to correspond to the quantum Fateev model \cite{bazhanov2014integrable}.
A generalization to affine Lie superalgebras \cite{zeitlin2014superopers} would also be interesting to study to explore the integrable structure of superstring theory in AdS space-time.

\subsection*{Note added:}
During the preparation of this paper, we became aware of \cite{Masoero:2015lga} where the conformal limit has been also studied for simply-laced cases.

\subsection*{Acknowledgments}

We would like to thank J. Suzuki for useful discussions.
The work of KI is supported in part by the JSPS Japan-Hungary Research Cooperative Program.

\appendix
\section{$T$-$Q$ relations for $A_{2r}^{(2)}$} \label{app:A2r_TQ}

In this appendix we will derive the Bethe ansatz equations \eqref{eq:A2r_BAE} for $A^{(2)}_{2r}$ starting from the ODE satisified by the top component of $\Psi^{(1)}$ in the conformal limit.
This was done for the $A_2^{(2)}$ case in \cite{Dorey:1999pv}.
For this discussion the $\psi$-system will not be used explicitly, but for reference we write down the $\psi$-system here where in the conformal limit it reduces to Wronskian relations on the top component of each vector $\Psi^{(a)}$,
\begin{equation}
  \begin{gathered}
    W[\psi^{(a)}_{-1/2} , \psi^{(a)}_{ 1/2}] = \psi^{(a-1)} \psi^{(a+1)} , \\
    W[\psi^{(r)}_{-1/2} , \psi^{(r)}_{ 1/2}] = \psi^{(r-1)} \psi^{(r)} .
  \end{gathered} \label{eq:psi_sys_a2r}
\end{equation}

In the conformal limit the top component of $\Psi^{(1)}$, $\psi^{(1)}$, satisfies the ODE (see \cite{Ito2014600})
\begin{gather}
  (\partial_x - \beta h^{(1)}_1 \cdot \partial_x \phi) \cdots (\partial_x - \beta h^{(1)}_r \cdot \partial_x \phi) \partial_x (\partial_x + \beta h^{(1)}_r \cdot \partial_x \phi) \cdots (\partial_x + \beta h^{(1)}_1 \cdot \partial_x \phi) \ \psi^{(1)} \notag\\
  = -p(x,E) \ \psi^{(1)} \,. \label{eq:A22_holo_eq}
\end{gather}
This equation has a subdominant solution $\psi$ with asymptotic behavior
\begin{gather}
  \psi \sim x^{-rM} \exp\left( -\frac{x^{M+1}}{M+1} \right) \,. \label{eq:psi_a2r}
\end{gather}
A Symanzik rotation of $\psi(x,E,g)$ is defined to be
\begin{gather}
  \psi_k(x,E,g) = \omega^{-kr} \psi(\omega^{k} x, \omega^{hMk} E, g) \,, \quad \omega = e^{2\pi i / h(M+1)} .
\end{gather}
Defining the Wronskian of $a$ such solutions to be
\begin{gather}
  W^{(a)}_{k_1, \ldots, k_{a}} = W^{(a)}[\psi_{k_1}, \ldots, \psi_{k_{a}}],
\end{gather}
then it follows that since the ODE \eqref{eq:A22_holo_eq} has no derivative of order $2r$,
\begin{gather}
  W^{(2r+1)}_{k,k+1,\ldots, k+2r} = \emph{const} \,.
\end{gather}
Substituting the asymptotics of \eqref{eq:psi_a2r} into the left side of this equation, the constant is shown to be nonzero and the solutions $\{ \psi_k, \psi_{k+1}, \ldots, \psi_{k+2r} \}$ are linearly independent.
We will also make use of the notation
\begin{gather}
  W^{(a)}_{k} = W^{(a)}[\psi_{k}, \psi_{k+1}, \ldots, \psi_{k+a-1}] ,
\end{gather}
and define the auxiliary functions which will make up the $\psi$-system
\begin{gather}
  \psi^{(a)} = W^{(a)}_{-\tfrac{a-1}{2}} . \label{eq:psia_a2r}
\end{gather}

To show that the above functions \eqref{eq:psia_a2r} asymptotically satisfy $\psi$-system \eqref{eq:psi_sys_a2r}, first note that the above $\psi_k$ asymptotic functions are exactly what one would get in the case of $A_{2r}^{(1)}$.
The work of \cite{dorey2007pseudo,juanjuan2012polynomial} then gives a $\psi$ system for auxiliary functions $\psi^{(1)}_{su(h)}, \ldots, \psi^{(2r)}_{su(h)}$.
The twisting of $A^{(1)}_{2r}$ to $A^{(2)}_{2r}$ implies that we expect $\psi^{(a)} = \psi^{(2r+1-a)}$.
Using the trigonometric relations one can show indeed that
\begin{gather}
  \psi^{(a)} \sim \psi^{(2r+1-a)} \sim x^{\left(-ra+\tfrac{a(a-1)}{2}\right)(M+1)} \exp\left( -\frac{x^{M+1}}{M+1} \frac{\sin(\pi a / h)}{\sin(\pi/h)} \right) .
\end{gather}
Notice that the coefficient in the exponential here is exactly $\mu^{(a)} / \mu^{(1)}$, as required.
This demonstrates that after making the identification $\psi^{(r+1)} \sim \psi^{(r)}$ the $\psi$-system of $A^{(1)}_{2r}$ reduces to \eqref{eq:A2r_psi}.
In the case of $A_{2r}^{(1)}$ one cannot truly identify $\psi^{(r+1)} \sim \psi^{(r)}$, but for $A^{(2)}_{2r}$ in addition to the large $x$ behavior the small $x$ behavior is also in agreement,
\begin{gather}
  W^{(r)} \sim W[\chi_1, \ldots, \chi_r] \sim x^{-\beta(h_1 + \cdots + h_r)\cdot g} \,, \\
  W^{(r+1)} \sim W[\chi_1, \ldots, \chi_r, \chi_{r+1}] \sim x^{-\beta(h_1 + \cdots + h_r)\cdot g} \,.
\end{gather}

Now, using $\psi$-system \eqref{eq:psi_sys_a2r} the Bethe ansatz equations \eqref{eq:A2r_BAE} can be proven to hold through the $T$-$Q$ relations we will now derive.
Since $\{ \psi_k, \psi_{k+1}, \ldots, \psi_{k+2r} \}$ form a basis of solutions, we can expand $\psi$ as
\begin{gather}
  \psi = \sum_{k=1}^{2r+1} C^{(k)} \psi_k \,.
\end{gather}
Then, using the notation $W^{(a)}_k = W^{(a)}_{k, k+1, \ldots, k+a-1}$ and determinant relations in \cite{dorey2000differential} (equation 4.5) gives
\begin{gather}
  T^{(1)}(E) \prod_{j=0}^{2r+1} W^{(j)}_0 = \sum_{m=0}^{2r} \left( \prod_{j=0}^{m-1} W^{(j)}_0 \right) W_1^{(m)} W_{-1}^{(m+1)} \left( \prod_{j=m+2}^{2r+1} W_0^{(j)} \right) \label{eq:CW_rel_final} ,
\end{gather}
where $T^{(1)}(E) \equiv C^{(1)}(\omega^{-hM} E)$ and the Coxeter number $h$ is $2r+1$ in this case.

We will also expand $\psi^{(a)}$ in terms of solutions defined by the small $x$ behavior as
\begin{gather}
  \psi^{(a)} = \sum_{i=1}^{\dim(V^{(a)})} Q^{(a)}_{[i]} \chi^{(a)}_{i} . \label{eq:psia_small_expand}
\end{gather}
After considering just the most divergent first term ($i=1$) in this expansion, we can then make the identification ($Q^{(a)}_{[1]} \equiv Q^{(a)}$)
\begin{gather}
  W^{(a)}_k \rightarrow \omega^{k\left[a\left(\tfrac{a-1}{2}-r\right)-\beta\omega_a \cdot g\right]} Q^{(a)}_{k+\tfrac{a-1}{2}} \,,\\
  W^{(2r+1-a)}_k \rightarrow \omega^{k\left[(2r+1-a)\left(\tfrac{2r-a}{2}-r\right)-\beta\omega_a \cdot g\right]} Q^{(a)}_{k+\tfrac{2r-a}{2}} \,.
\end{gather}
Therefore the $T$-$Q$ system becomes
\begin{gather}
T^{(1)}(E) \prod_{j=0}^{2r+1} Q^{(j)}_{(j-1)/2} = \sum_{m=0}^{2r} \left( \prod_{j=0}^{m-1} Q^{(j)}_{(j-1)/2} \right) \omega^{\beta_m - \beta_{m+1}} Q_{1+(m-1)/2}^{(m)} Q_{-1+m/2}^{(m+1)} \left( \prod_{j=m+2}^{2r+1} Q_{(j-1)/2}^{(j)} \right) \label{eq:TQ_rel} ,\\
  \beta_m = m\left(\tfrac{m-1}{2}-r\right)-\beta\omega_m \cdot g , \quad \beta_0 = \beta_{2r+1} = 1 , \quad \omega_{2r+1-a} = \omega_{a} ,\\
  Q^{(a)}_k = Q^{(2r+1-a)}_k , \quad Q^{(0)} = Q^{(2r+1)} = 1 .
\end{gather}
Evaluating the $T$-$Q$ system at $E = \omega^{hM(a-1)/2} E^{(a)}_i$ where $Q^{(a)}(E^{(a)}_i) = 0$ then gives equations \eqref{eq:A2r_BAE} as desired.

\section{Comparison with Kuniba-Suzuki for $A^{(2)}_{2r}$} \label{app:kuniba_suzuki}

In this appendix we show how the Bethe ansatz equations for $A^{(2)}_{2r}$ \eqref{eq:A2r_BAE} are in agreement with \cite{kuniba_suzuki_1995}, which looked at Bethe ansatz equations for twisted quantum affine algebras (see also \cite{reshetikhin1987towards}).

The Bethe ansatz equation associated to a solvable vertex model associated with the twisted quantum affine algbra $U_q(A^{(2)}_{2r})$ \cite{kuniba_suzuki_1995} is
\begin{gather}
  - \prod_{t=0}^1 \frac{\phi(i u_j^{(a)} + \langle s \omega_{\dot{p}} | \alpha_{\sigma^t(\dot{a})}\rangle + \tfrac{t\pi i}{2\hbar})}{\phi(i u_j^{(a)} - \langle s \omega_{\dot{p}} | \alpha_{\sigma^t(\dot{a})}\rangle + \tfrac{t\pi i}{2\hbar})} = \prod_{t=0}^1 \prod_{b \in \hat{S}} \frac{Q^{(b)}(i u^{(a)}_j + \langle \alpha_{\dot{a}} | \alpha_{\sigma^t(\dot{b})} \rangle + \tfrac{t \pi i}{2 \hbar})}{Q^{(b)}(i u^{(a)}_j - \langle \alpha_{\dot{a}} | \alpha_{\sigma^t(\dot{b})} \rangle + \tfrac{t \pi i}{2 \hbar})} \,,
\end{gather}
where
\begin{gather}
  [u]_k = q^{ku} - q^{-ku} \,, \quad q = e^\hbar, \\
  \phi(u) = \prod_{j=1}^N [u-w_j]_1 \,, \quad Q^{(a)} = \prod_{j=1}^{N^{(a)}} [u - i u^{(a)}_j]_1 \,.
\end{gather}
Note that the effect of the twist term is such that
\begin{gather}
  [u + \tfrac{t \pi i}{2 \hbar}]_1 = \left\{ \
    \begin{matrix}
      2 \sinh( \hbar u ) \quad \text{for} \; t=0 \\
      2i \cosh( \hbar u) \quad \text{for} \; t=1
    \end{matrix} \right. \,.
\end{gather}

Define $\hbar i u_j^{(a)} = v_j^{(a)}$, and set $2\hbar = -i \theta$.
For large $N$ in the thermodynamic limit, we have $v \rightarrow \infty$.
We define the finite part of $v_i^{(a)}$ as $x_i^{(a)}$ through the equation
\begin{gather}
  v_j^{(a)} \equiv \frac{1}{2\rho} (x_j^{(a)} + \log N) \,. \label{eq:thermolimit}
\end{gather}
Then, defining $E^{(a)}_i \equiv \exp\big(\tfrac{x^{(a)}_i}{\rho}\big)$ we get
\begin{gather}
  - e^{-iN\theta s (\delta_{\dot{a},\dot{p}} + \delta_{\sigma(\dot{a}),\dot{p}}) -iN^{(a)} \theta \langle \alpha_{\dot{a}} | \sum_b \alpha_{\dot{b}} + \alpha_{\sigma(\dot{b})} \rangle} = \quad\quad\quad\quad\quad \notag\\
  \quad\quad\quad\quad\quad \prod_{b \in \hat{S}} \prod_{k=1}^{N^{(b)}}
  \frac{
    (E^{(a)}_j e^{-i\theta \langle \alpha_{\dot{a}} | \alpha_{\dot{b}} \rangle} - E^{(b)}_k)
    (E^{(a)}_j e^{-i\theta \langle \alpha_{\dot{a}} | \alpha_{\sigma(\dot{b})} \rangle} + E^{(b)}_k)
  }{
    (E^{(a)}_j e^{i\theta \langle \alpha_{\dot{a}} | \alpha_{\dot{b}} \rangle} - E^{(b)}_k)
    (E^{(a)}_j e^{i\theta \langle \alpha_{\dot{a}} | \alpha_{\sigma(\dot{b})} \rangle} + E^{(b)}_k)
  } \,.
\end{gather}
Finally, making the definition $Q^{(a)}(E) \equiv \prod_{k=1}^{N^{(a)}} (E - E^{(a)}_k)$,
the BAE becomes
\begin{gather}
  - e^{-iN\theta s (\delta_{\dot{a},\dot{p}} + \delta_{\sigma(\dot{a}),\dot{p}}) -iN^{(a)} \theta \langle \alpha_{\dot{a}} | \sum_b \alpha_{\dot{b}} + \alpha_{\sigma(\dot{b})} \rangle} = \quad\quad\quad\quad\quad \notag\\
  \quad\quad\quad\quad\quad \prod_{b \in \hat{S}}
  \frac{
    Q^{(b)}(E^{(a)}_j e^{-i\theta \langle \alpha_{\dot{a}} | \alpha_{\dot{b}} \rangle})
    Q^{(b)}(-E^{(a)}_j e^{-i\theta \langle \alpha_{\sigma(\dot{a})} | \alpha_{\dot{b}} \rangle})
  }{
    Q^{(b)}(E^{(a)}_j e^{i\theta \langle \alpha_{\dot{a}} | \alpha_{\dot{b}} \rangle})
    Q^{(b)}(-E^{(a)}_j e^{i\theta \langle \alpha_{\sigma(\dot{a})} | \alpha_{\dot{b}} \rangle})
  } \,. \label{eq:general_BAE}
\end{gather}

In the $A_{2r}^{(2)}$ case, for these Bethe ansatz equations to agree with \eqref{eq:A2r_BAE}, simultaneously replace $E_j^{(a)} \rightarrow -E_j^{(a)}$ for odd $a$, set $\tilde{\theta} = \pi - \theta$ and take $N$ and $N^{(1)}$ to be even.
The identification then holds for $Q^{(a)}_k(E) = Q^{(a)}(\omega^{hMk}E)$ where $\omega = e^{2\pi i / h(M+1)}$ when $\tilde{\theta} = \frac{\pi M}{M+1}$.
%

\end{document}